# Nuclear-fission studies with relativistic secondary beams: analysis of fission channels

C. Böckstiegel[a], S. Steinhäuser[a], K.-H. Schmidt[b,1], H.-G. Clerc[a], A. Grewe[a], A. Heinz[a,2], M. de Jong[a], A. R. Junghans[b,3], J. Müller[a], B. Voss[b]

[a]*Institut für Kernphysik, Technische Universität Darmstadt, Schloßgartenstraße 9, 64289 Darmstadt, Germany*

[b]*Gesellschaft für Schwerionenforschung m. b. H., Planckstraße 1, 64291 Darmstadt, Germany*

**Abstract.** Nuclear fission of several neutron-deficient actinides and pre-actinides from excitation energies around 11 MeV was studied at GSI Darmstadt by use of relativistic secondary beams. The characteristics of multimodal fission of nuclei around $^{226}$Th are systematically investigated and interpreted as the superposition of three fission channels. Properties of these fission channels have been determined for 15 systems. A global view on the properties of fission channels including previous results is presented. The positions of the asymmetric fission channels are found to be constant in element number over the whole range of systems investigated.

## INTRODUCTION

In a previous publication [1] we have reported on a systematic experimental study of fission-fragment element yields and kinetic energies in the fission of 70 neutron-deficient actinides and pre-actinides between $^{205}$At and $^{234}$U. In the present work, the data of this experiment are interpreted within the concept of independent fission channels [2, 3] according to which the fissioning system follows specific valleys in the potential energy in the direction of elongation. The fission channels are characterized by several parameters, e.g. the average mass or charge split, the mass or charge width, and the mean total kinetic energy, respectively the elongation of the scission configuration. We extract the values of these parameters for three fission channels from the measured

[1] Corresponding author. E-mail address: k.h.schmidt@gsi.de, URL: www.gsi.de/charms
[2] Present address: *Wright Nuclear Structure Laboratory, Yale University, New Haven, CT 06520, USA*
[3] Present address; *Forschungszentrum Rossendorf, Postfach 510119, 01314 Dresden, Germany*

data of 15 of the systems, which show features of multi-modal fission. The systematic survey of fissioning systems in the transition from single-humped to double-humped element distributions around $^{226}$Th extends the systematic view on how the intensities and other relevant parameters of the fission channels vary as a function of the nuclear composition of the fissioning nucleus.

## EXPERIMENT

At GSI Darmstadt, a new technique to investigate low-energy fission has been developed [4, 5]. Relativistic secondary projectiles are produced via fragmentation of a 1 *A* GeV primary beam of $^{238}$U and identified in nuclear mass and charge number by the fragment separator FRS [6]. In a dedicated experimental set-up, the giant resonances, mostly the giant dipole resonance, are excited by electromagnetic interactions in a secondary lead target, and fission from excitation energies around 11 MeV is induced. The fission fragments are identified in nuclear charge, and their velocity vectors are determined. From these data, the element yields and the total kinetic energies are deduced. Details of the experimental technique are given elsewhere [1].

## RESULTS

The measured element-yield distributions and the total kinetic energies of $^{233}$U, $^{232}$Pa, $^{228}$Pa, $^{228}$Th, $^{226}$Th, and $^{223}$Th are shown in Figure 1. The gross structural effects observed in the element yields are different from those showing up in the kinetic energies. From $^{233}$U to $^{223}$Th, the weight of the symmetric fission component in the element yields increases strongly to the expense of the asymmetric component. In contrast, the general features of the total kinetic energies are preserved. These are enhanced kinetic-energy values near Z = 52 to 54 and reduced values at symmetry. In a simultaneous fit to elemental yields and total kinetic energies, it was possible to reproduce these data with the assumption of independent fission channels.

A satisfactory description was obtained with three channels, "standard I" close to $N$ = 82, "standard II" around $N$ = 88 in the heavy fragment, and "super long" at symmetry, according to the notations introduced by Brosa et al. [3]. Each channel was represented by a Gaussian distribution in the yields and a specific elongation of the scission-point configuration. In order to consider the trivial variation of the total kinetic energy as a function of mass and charge split, the Coulomb repulsion $V_C$ in the scission-point configuration was parametrised by the following expression, introduced in ref. [7]:

$$V_C = \frac{Z_1 \cdot Z_2 \cdot e^2}{r_0 \left( A_1^{1/3} \cdot \left[ 1 + \frac{2\beta_1}{3} \right] + A_2^{1/3} \cdot \left[ 1 + \frac{2\beta_2}{3} \right] \right) + d} \qquad (1)$$

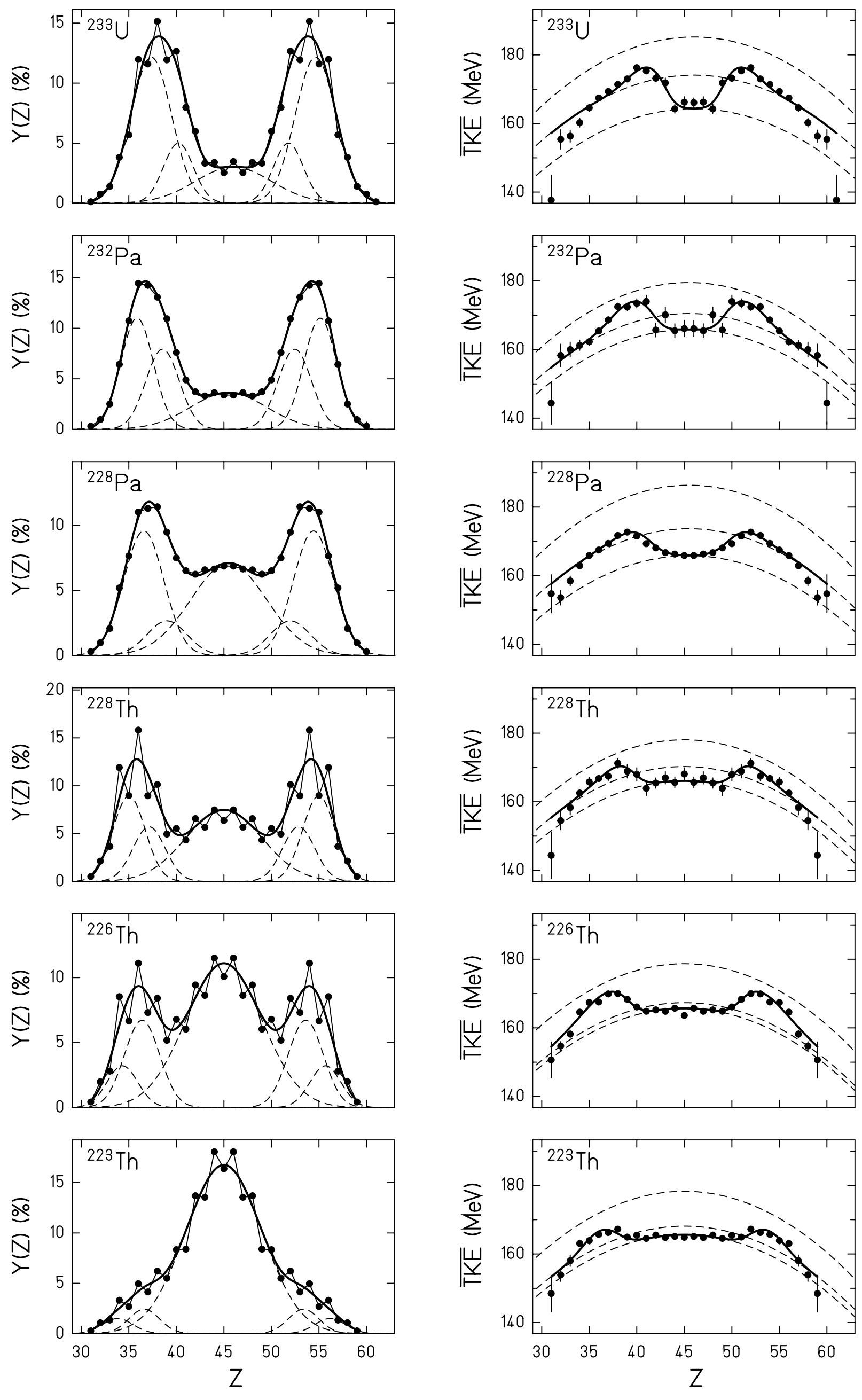

**Figure 1.** The data points mark the measured elemental yields (left column) and average total kinetic energies (right column) as a function of the nuclear charge of the fission fragments for several fissioning nuclei. Only statistical errors are given. They are not shown when they are smaller than the symbols. The total kinetic energies are subject to an additional systematic uncertainty of 2 %, common to all data. The full lines show descriptions with the model of independent fission channels with the parameters given in table 1. Dashed lines depict the contributions of the individual channels. The sequence from symmetry to largest asymmetry in the yields is: super long, standard I, standard II. The sequence from lowest to highest *TKE* is: super long, standard II, standard I. (See text for details.)

**Table 1.** Parameters of the independent fission channels determined in this work. The positions in *Z* are given for the heavy group. The width corresponds to the standard deviation in *Z*. The width of the super-long channel was fixed to 4.0 charge units from the systematics of lighter systems established in ref. [1].

| Nucleus | Channel | *d* / fm | Position | Width | Yield (%) |
|---|---|---|---|---|---|
| | Standard I | 0.88 ± 0.25 | 52.7 ± 0.6 | 2.13 ± 0.22 | 49.7 ± 20.6 |
| $^{234}$U | Standard II | 2.06 ± 0.47 | 55.2 ± 0.5 | 2.07 ± 0.14 | 39.1 ± 20.1 |
| | Super long | 2.83 ± 0.32 | 46 | 4.0 | 11.2 ± 1.1 |
| | Standard I | 0.59 ± 0.21 | 51.8 ± 0.2 | 1.57 ± 0.12 | 20.0 ± 5.5 |
| $^{233}$U | Standard II | 1.63 ± 0.06 | 54.6 ± 0.2 | 2.14 ± 0.06 | 66.2 ± 5.4 |
| | Super long | 2.68 ± 0.13 | 46 | 4.0 | 13.8 ± 0.5 |
| | Standard I | 0.85 ± 0.15 | 52.4 ± 0.2 | 1.91 ± 0.07 | 35.2 ± 3.4 |
| $^{232}$U* | Standard II | 1.69 ± 0.10 | 55.1 ± 0.2 | 1.91 ± 0.07 | 46.9 ± 3.9 |
| | Super long | 2.19 ± 0.18 | 46 | 4.0 | 17.9 ± 0.8 |
| | Standard I | 0.77 ± 0.16 | 52.5 ± 0.4 | 1.75 ± 0.17 | 34.9 ± 9.8 |
| $^{232}$Pa | Standard II | 1.64 ± 0.13 | 55.1 ± 0.2 | 1.70 ± 0.08 | 47.0 ± 9.6 |
| | Super long | 2.13 ± 0.16 | 45.5 | 4.0 | 18.1 ± 0.8 |
| | Standard I | 0.74 ± 0.10 | 52.5 ± 0.1 | 1.85 ± 0.04 | 31.8 ± 2.5 |
| $^{231}$Pa* | Standard II | 1.54 ± 0.05 | 55.0 ± 0.1 | 1.85 ± 0.04 | 48.8 ± 2.7 |
| | Super long | 2.24 ± 0.09 | 45.5 | 4.0 | 19.4 ± 0.4 |
| | Standard I | 0.35 ± 0.20 | 52.6 ± 0.2 | 1.95 ± 0.05 | 26.9 ± 4.3 |
| $^{230}$Pa* | Standard II | 1.60 ± 0.08 | 54.9 ± 0.1 | 1.95 ± 0.05 | 50.6 ± 4.5 |
| | Super long | 2.29 ± 0.09 | 45.5 | 4.0 | 22.5 ± 0.5 |
| | Standard I | 0.45 ± 0.13 | 52.5 ± 0.1 | 1.87 ± 0.05 | 27.6 ± 2.8 |
| $^{229}$Pa* | Standard II | 1.57 ± 0.07 | 54.9 ± 0.1 | 1.87 ± 0.05 | 47.0 ± 3.0 |
| | Super long | 2.38 ± 0.08 | 45.5 | 4.0 | 25.4 ± 0.5 |
| | Standard I | 0.25 ± 0.30 | 52.0 ± 0.3 | 2.11 ± 0.04 | 13.8 ± 4.0 |
| $^{228}$Pa* | Standard II | 1.42 ± 0.06 | 54.4 ± 0.1 | 2.11 ± 0.04 | 50.7 ± 4.2 |
| | Super long | 2.23 ± 0.05 | 45.5 | 4.0 | 35.5 ± 0.4 |
| | Standard I | 0.63 ± 0.24 | 52.9 ± 0.3 | 1.70 ± 0.09 | 24.4 ± 5.7 |
| $^{228}$Th* | Standard II | 1.38 ± 0.12 | 55.0 ± 0.2 | 1.70 ± 0.09 | 38.1 ± 6.0 |
| | Super long | 1.81 ± 0.10 | 45 | 4.0 | 37.5 ± 1.0 |
| | Standard I | 0.62 ± 0.10 | 53.6 ± 0.3 | 1.85 ± 0.13 | 31.2 ± 7.4 |
| $^{226}$Th | Standard II | 1.72 ± 0.52 | 55.7 ± 0.3 | 1.63 ± 0.10 | 13.2 ± 7.3 |
| | Super long | 1.90 ± 0.03 | 45 | 4.0 | 55.6 ± 0.4 |
| | Standard I | 0.82 ± 0.12 | 53.8 ± 0.2 | 1.85 ± 0.08 | 25.4 ± 3.4 |
| $^{225}$Th | Standard II | 2.44 ± 0.50 | 56.1 ± 0.5 | 1.82 ± 0.15 | 6.0 ± 3.3 |
| | Super long | 2.01 ± 0.02 | 45 | 4.0 | 68.6 ± 0.4 |
| | Standard I | 0.18 ± 0.47 | 52.8 ± 0.3 | 1.51 ± 0.16 | 6.3 ± 3.2 |
| $^{224}$Th | Standard II | 1.48 ± 0.10 | 55.0 ± 0.3 | 1.93 ± 0.09 | 17.9 ± 3.2 |
| | Super long | 1.98 ± 0.02 | 45 | 4.0 | 75.8 ± 0.4 |
| | Standard I | 0.72 ± 0.11 | 53.4 ± 0.1 | 1.61 ± 0.06 | 9.9 ± 0.4 |
| $^{223}$Th* | Standard II | 1.71 ± 0.15 | 56.2 ± 0.1 | 1.61 ± 0.06 | 6.1 ± 0.4 |
| | Super long | 1.98 ± 0.02 | 45 | 4.0 | 84.0 ± 0.4 |
| | Standard I | 0.85 ± 0.20 | 53.3 ± 0.2 | 1.65 ± 0.11 | 6.6 ± 0.6 |
| $^{222}$Th* | Standard II | 1.64 ± 0.19 | 56.0 ± 0.2 | 1.65 ± 0.11 | 5.2 ± 0.7 |
| | Super long | 1.97 ± 0.02 | 45 | 4.0 | 88.2 ± 0.6 |
| | Standard I | 0.85 ± 0.35 | 52.2 ± 0.2 | 0.84 ± 0.30 | 3.8 ± 2.6 |
| $^{223}$Ac | Standard II | 0.90 ± 0.22 | 54.5 ± 0.7 | 1.89 ± 0.40 | 7.8 ± 2.6 |
| | Super long | 1.65 ± 0.03 | 44.5 | 4.0 | 88.4 ± 1.2 |

*) For these nuclei the widths of standard I and standard II were set equal in the fit.

$Z_i$, $A_i$ and $\beta_i$ are nuclear-charge numbers, mass numbers and quadrupole deformations of the fission fragments, $r_0$ = 1.16 fm is the nuclear-radius constant, and $e$ the elementary charge. The mass numbers were related to the charge numbers by the unchanged-charge-density (UCD) assumption; this means that the neutron-to-proton ratio of the fission fragments is assumed to be equal to that of the fissioning nucleus. Evaporation was neglected. The deformation parameters were fixed at $\beta_i$ = 0.6 as predicted by the liquid-drop model, see ref. [7]. The "tip distance" *d* was determined from a fitting procedure, requiring that $V_C$ best reproduces the measured *TKE* values.

Please note that the variation of the tip distance *d* effectively accounts for the variations of all three parameters, $\beta_1$, $\beta_2$ and *d,* but the experimental information is not sufficient to determine the values of $\beta_1$, $\beta_2$ and *d* individually. It is expected that the deformation parameters vary most strongly from one fission channel to another.

For each fission channel, position, width, and area of the Gaussian representing the nuclear-charge yields as well as the tip distance of the scission configuration were treated as free parameters. The width of the super-long channel had to be kept constant for the fit to converge. The value was taken from the systematics of lighter systems determined in ref. [1]. The yields are formulated as the sum, the total kinetic energies as the weighted average of the different components. The results of the fit are given in Table 1. Unfortunately, the dispersion of the total kinetic energy could not be determined in the secondary-beam experiment due to the limited energy resolution. Therefore, the relative weights of the two asymmetric fission channels could only be determined with rather large error bars.

## DISCUSSION

The parameters, in particular the magnitudes of the total kinetic energies, attributed to the individual fission channels, roughly coincide with the expectations from systematics of the three most intense fission channels known from heavier fissioning systems, see e.g. refs. [8, 9, 10, 11, 12, 13, 14, 15, 16, 17]. Also for thorium isotopes produced at higher excitation energies by fusion reactions and for nuclei in the lead region, standard I and standard II fission channels have been observed [18, 19, 20]. As a remarkable result, we found that the effective tip distance of the super-long channel, which appears at symmetry, becomes smaller, and hence the scission configuration of this channel becomes more compact for the lighter systems. This finding indicates that shell effects influence this channel, too, although the charge distributions measured in the present experiment can be represented by simple Gaussians. There is no indication of a fine structure in the shape of the symmetric fission channel, which was found for nuclides from $^{187}$Ir to $^{213}$Ac in ref. [21]., The element distributions show only an even-odd staggering due to pairing correlations [22]. The variation of the tip distance in the symmetric channel found in the present work may be related to the properties of the broad deformed shell around $N$ = 64, which tends to become less deformed with decreasing neutron number (see e. g. refs. [7, 23]).

From the good simultaneous description of nuclear-charge yields and total kinetic energies as demonstrated in Figure 1 we conclude that the concept of independent fission channels is well suited to describe the present data. It allows for the strong variations of the yields while keeping the *TKE* distributions almost unchanged. However, we would also like to add a critical remark. In accordance with the usual treatment, the most probable scission-point configuration, parameterised by the tip distance $d$, was fixed at one value for each fission channel and not allowed to vary as a function of charge split within a given fission channel. This might be an oversimplification. It is known that the energetically most favourable deformation in a deformed shell region varies with the number of nucleons [7, 23]. Therefore, a corresponding variation of the mean elongation at the scission point as a function of the charge split is to be expected. This would imply a variation of the most probable effective tip distance $d$ even inside one fission channel. Besides the symmetric channel, this also concerns the standard II fission channel related to the $N \approx 88$ deformed shell. Therefore, part of the increase of the *TKE* values in the asymmetric component towards symmetry, which is attributed to the compact standard I channel in our description, may rather be attributed to a variation of the scission configuration in the standard II channel to more compact shapes. This would affect all parameters of the two asymmetric fission channels.

The data on elemental yields seem to support the idea, stated by Itkis et al. [19], that the weights of the fission channels are principally determined by an interplay of the neutron shells at $N = 82$ and $N \approx 88$ with the liquid-drop potential. The total kinetic energies, however, seem to be closely related to the structural properties of the fission fragments. This finding agrees with the expectation that the shell effects in the scission-point configuration essentially determine the total kinetic energies.

In Figure 2, we compare the parameters of the independent fission channels determined in the present work with the body of previously available data. This allows a systematic view on the variation of position and width of the standard I and standard II channels, on the width of the super-long channel and on the relative yields of these three channels as a function of element number and mass number of the fissioning system. The data are restricted to spontaneous fission and to initial excitation energies up to a few MeV above the fission barrier, where structural effects are expected to be strong. At the first glance it is not obvious how to deduce global trends from these data, since they are subject to large fluctuations. Unfortunately, not all parameter values of the fission channels determined in previous work have been published, and part of the data is only given without specifying any uncertainty range. This makes a global analysis even harder. A list of the properties of the fissioning systems behind these data is given in Appendix 1. We would like to discuss the different parameters, one by one.

The widths of the standard I and standard II fission channels show fluctuations, which are appreciably larger than the reported statistical uncertainties of the fits. As figure 1 shows, these two fission channels overlap strongly in mass number. Therefore, we interpret the strong fluctuations of their widths as an indication for the strong correlation of these values in the fit. It seems that the reported statistical uncertainties are not realistic. No systematic trend can safely be deduced over the whole mass range. One may only conclude that the width of the standard I fission channel amounts to about 3.5 mass units, while the width of the standard II fission channel is appreciably larger with about 5 mass units.

The width of the super-long fission channel is rather well determined for the light systems of the present experiment, for which the symmetric fission component dominates, to about 10 mass units over a large mass range, see ref. [1]. Thus, the super-long fission channel is much broader than the standard I and standard II fission channels. The values deduced previously for heavier systems fluctuate enormously. These fluctuations are probably explained by the tiny yield of the super-long channel for the heavier systems. If the super-long channel could be observed at all, only its central part was seen. The wings, which are necessary to determine the width, were completely covered by the standard I and standard II fission channels. One is tempted to assume that the width of the super-long fission channel follows the trend of the lighter systems over the whole mass range: The width is about 10 mass units with a slight trend to smaller values with increasing mass.

The positions of the standard I and standard II fission channels show a systematic variation as a function of mass number for a given element. This trend is clearly seen for all elements, in spite of some fluctuations. The fluctuations are roughly consistent with the magnitude of the statistical uncertainties of the fit, depicted by the error bars. The average slope of the isotopic trend seems to be slightly larger than 0.5. This means that the positions of the standard I and standard II fission channels vary in neutron number and are rather stable in proton number. This surprising result has already been observed for the position of the lumped asymmetric component, the total yield of standard I and standard II, see Figure 22 of ref. [1]. It seems now that this feature extends over the whole range of elements where such data are available, that means up to americium.

The stability of the positions of the two standard fission channels in element number is even more evident in Figure 3. For most of the systems, the mean nuclear charge of the standard I fission channel is close to 52.5, while the mean nuclear charge of the standard II fission channel is close to 55, with no clear systematic variation as a function of element number or neutron excess. Correspondingly, the variation of neutron excess for fissioning systems in an isotopic chain of a given element is reflected by a strong variation of the position of the two standard channels in neutron number as demonstrated in the lower part of Figure 3. This finding sheds a new light on the well known observation of Unik et al. [24], who stated that the position of the heavy component of the fission-fragment distribution in asymmetric fission is approximately constant in mass over the whole range of fissioning systems investigated. On the basis of Figures 2 and 3, we re-formulate more precisely: It is not the mass number but the element number, which is primarily kept constant. The variation of the mean mass number of the heavy component for the measured systems remains relatively small only due to the limited $N/Z$ range of the fissioning systems, which could be investigated up to now. It is beyond the scope of the present work to discuss the theoretical implications of this finding, but we would like to stress that it is quite astonishing, since shell-model calculations suggest that the neutron shells are generally stronger than the proton shells, and thus neutron shells are assumed to have a dominant influence on the fission process compared to proton shells. Therefore, shell-model calculations rather suggest that the positions of the two standard fission channels should be stable in neutron number.

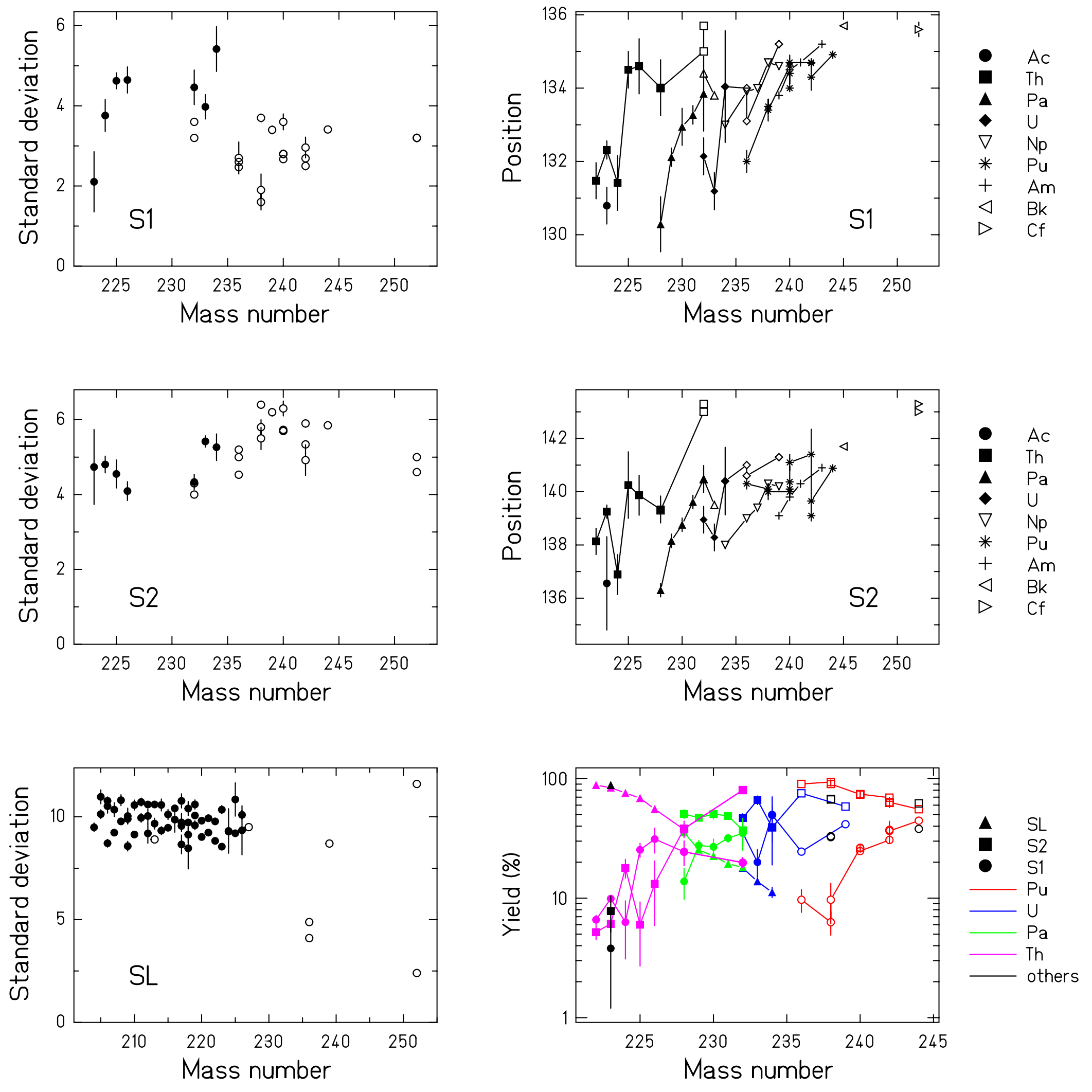

**Figure 2:** (Colour online) Global view on the parameters of the three independent fission channels standard I (S1), standard II (S2) and super long (SL). The results of the present work (full symbols), supplemented by some values for the super-long fission channel from ref. [1] (also marked by full symbols), are compared with other available experimental data (open symbols) [3, 9, 11, 12, 13, 15, 16, 25, 26, 27, 28, 29, 30, 31]. All data are given in mass numbers. Values measured in nuclear charge were converted to mass numbers using the unchanged-charge-density assumption and neglecting neutron evaporation.

The last characteristic of Figure 2 to discuss is the variation of the relative yields of the three fission channels. There is a clear tendency to be observed at the first glance: The relative yield of the symmetric super-long fission channel shows an exponential de-crease with increasing mass number. For systems with $A > 234$, the yield of the super-long channel becomes so low that it could not be determined any more. At the same time, the complementary yield of the lumped asymmetric component increases with increasing mass. However, any systematic trend in the competition between the standard I and the standard II fission channels is less obvious.

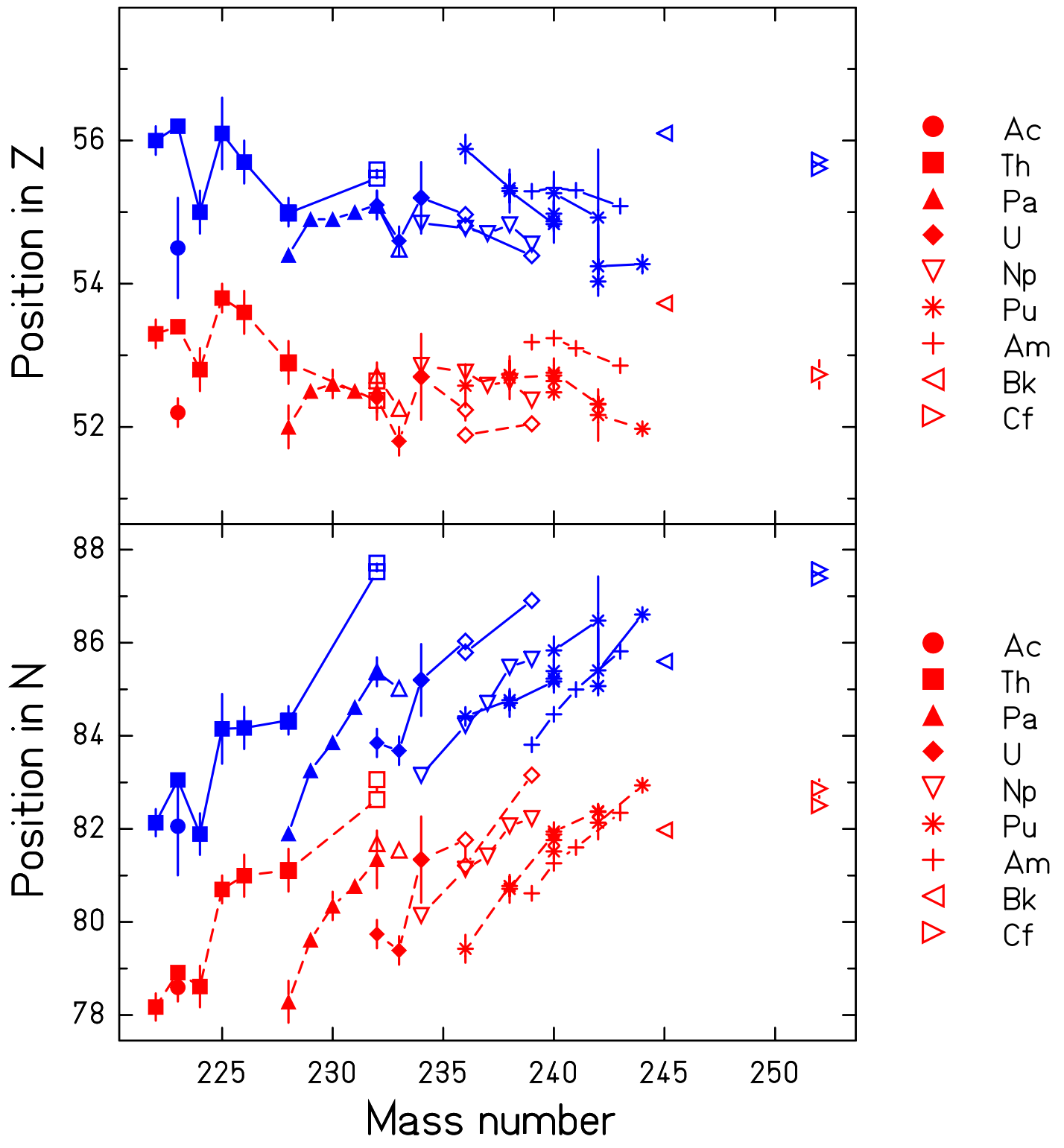

**Figure 3:** (Colour online) Mean positions of the standard fission channels in element number (upper part) and neutron number (lower part) deduced from the data in Figure 2. Values were converted from measured element numbers or mass numbers using the unchanged-charge-density assumption and neglecting neutron evaporation. The values for the isotopes of a given element are connected by dashed (standard I) and full (standard II) lines. Data from the present experiment are marked by full symbols, those from other experiments by open symbols. The shape of the symbol denotes the element.

Analyzing the data in more detail, we would like to stress that only two elements have been investigated under good conditions and with good statistics over a long mass range. These are protactinium and plutonium. In the light thorium isotopes, the asymmetric component is rather weak, showing up only as shoulders on the wings of the super-long channel. This implies large uncertainties of the fit result on the weights of the standard I and the standard II fission channels. In light uranium isotopes, the statistics of our experiment is rather poor. Thus, we have a good justification to restrict our analysis to protactinium and plutonium. Interesting enough, these two elements show a consis-tent trend: The relative weight of the standard I fission channel grows strongly with increasing mass number. It is tempting to interpret this finding in the following way: With increasing mass, the $N/Z$ ratio of the fissioning system increases and comes closer to the value of $^{132}$Sn. If we relate the standard I fission channel to the common influence of the $Z = 50$ and $N = 82$ shells in the heavy fission fragment [32], it is reasonable that this influence increases if the nuclides produced in the standard I fission channel move closer to $^{132}$Sn on the chart of the nuclides.

The global analysis of the characteristics of the fission channels revealed interesting trends and features, which were not at all obvious from a restricted view on part of the data. The contribution from the present experiment has considerably enriched our empirical knowledge. The body of data presently available is consistent with the assumption that the parameters of the independent fission channels vary in a smooth and consistent way. However, the quality of the experimental results is far from being satisfactory, and the extension of the systems investigated in atomic number and in particular in neutron excess is still rather restricted.

## CONCLUSION

Elemental yields and total kinetic energies after low-energy fission were measured for a series of short-lived radioactive nuclei. The data nicely demonstrate the decisive influence of nuclear structure on the fission process in a particular interesting transitional region around $^{226}$Th. In contrast to the total kinetic energies, the element distributions were found to vary strongly, essentially as a function of mass number of the fissioning system. The model of independent fission channels could well account for this behaviour. The weights of the asymmetric fission channels decrease with decreasing mass of the fissioning nucleus, but the scission-point configurations remain essentially unchanged. The scission-point configuration of the symmetric fission channel, however, evolves to more compact shapes for the lighter fissioning nuclei, where symmetric splits correspond to smaller nucleon numbers. This clearly reveals the influence of shell effects also in the symmetric channel. In spite of a large scattering of great part of the data, a global analysis of the properties of fission channels revealed interesting global trends. The most salient feature is that the positions of the heavy components of the asymmetric fission channels do not vary in element number, while they move strongly in mass as well as in neutron number. The data are compatible with the assumption that the parameters of the fission channels vary in a smooth and systematic way between actinium and californium.

## ACKNOWLEDGEMENT

This work has been supported by the GSI Hochschulprogramm and by the Bundesministerium für Bildung, Wissenschaft, Forschung und Technologie under contract number BMBF 06 DA 473. The responsibility rests with the authors.

# APPENDIX 1: Compilation on properties of fission channels

Table A1 lists experimental results and some related calculated parameters on the properties of independent fission channels published previously. The results from each experiment performed at the lowest excitation energy are included in Figure 2. The table gives detailed information on the conditions of the experiment and on the result of the fit. Unfortunately, in many cases the numerical values of the fit have only partially been published and uncertainties are not always specified.

**Table A1.** Compilation on experimental and some related calculated properties of fission channels. In some cases, more than the standard I (S I), standard II (S II) and super-long (SL) fission channels are considered, like the super-asymmetric (SA) channel or the standard III (S III) channel. Please refer to the original publications for details.

| Ref. | Nucleus | Channel | Position | Width | Yield (%) | Remarks |
|---|---|---|---|---|---|---|
| BrG90 [3] | $^{213}$At | Standard | 137 | 3.9 | - | Experiment (See [3] for original work) |
| | | SL | 107 | 8.9 | - | |
| | $^{227}$Ac | Standard | 139 | 6.0 | - | |
| | | SL | 114 | 9.5 | - | |
| | $^{232}$Th | S I | 135 | 3.6 | - | |
| | | S II | 143 | 4.3 | - | |
| | $^{236}$U | S I | 134 | 2.6 | - | |
| | | S II | 141 | 5.0 | - | |
| | | SL | 118 | 4.1 | - | |
| | $^{240}$Pu | S I | 134 | 2.8 | - | |
| | | S II | 140 | 5.7 | - | |
| | $^{252}$Cf (s.f.) | S I | 135 | 3.2 | - | |
| | | S II | 143 | 5.0 | - | |
| | | S III | 149 | 7.1 | - | |
| | | S A | 178 | 2.3 | - | |
| | | SL | 127 | 11.6 | - | |

| Ref. | Nucleus | Channel | Position | Width | Yield (%) | Remarks |
|---|---|---|---|---|---|---|
| ScW92 [9] | $^{238}$Pu(s.f.) | S I | 133.4±0.3 | 1.9±0.4 | 9.7±3.6 | Experiment |
| | | S II | 140.0±0.3 | 5.5±0.3 | 90.3±5.5 | |
| | $^{240}$Pu(s.f.) | S I | 134.4±0.1 | 2.8±0.1 | 26.2±2.0 | |
| | | S II | 140.1±0.2 | 5.7±0.1 | 73.8±2.4 | |
| | $^{242}$Pu(s.f.) | S I | 134.7±0.1 | 2.5±0.1 | 30.7±1.8 | |
| | | S II | 139.1±0.2 | 5.9±0.1 | 69.3±2.2 | |
| | $^{239}$Pu($n_{th}$,f) | S I | 134.7±0.2 | 3.6±0.2 | 24.8±0.3 | |
| | | S II | 141.1±0.3 | 6.3±0.2 | 74.2±0.3 | |
| | | S III | 156.7±1.0 | 3.4±0.5 | 1.0±1.0 | |

| Ref. | Nucleus | Channel | Position | Width | Yield (%) | Remarks |
|---|---|---|---|---|---|---|
| PiJ93 [25] | $^{232}$Th | S I | 135.7 | 3.2 | 19.8 | Experiment |
| | | S II | 143.3 | 4.0 | 80.2 | $E_{end\text{-}point}$ = |
| | Brems- | SL | 116 | - | 0.0 | 6.44 MeV |
| | strahlung | S I | 135.2 | 3.6 | 30.0 | Experiment |
| | | S II | 142.7 | 4.3 | 68.8 | $E_{end\text{-}point}$ = |
| | | SL | 116 | - | 1.2 | 13.15 MeV |

| Ref. | Nucleus | Channel | Position | Width | Yield (%) | Remarks |
|---|---|---|---|---|---|---|
| AaW94 [26] | $^{252}$Cf (s.f.) | S I | 135.6±0.2 | 3.2±0.1 | 13.5±0.5 | Experiment |
| | | S II | 143.3±0.1 | 4.6±0.1 | 48.2±1.1 | |
| | | S III | 146.3±0.3 | 8.2±0.1 | 35.0±1.2 | |
| | | S A | 177.0±1.2 | 5.0±1.2 | 0.3±0.1 | |
| | | SL | 130.7±0.2 | 2.4±0.1 | 3.0±0.2 | |

| Ref. | Nucleus | Channel | Position | Width | Yield (%) | Remarks |
|---|---|---|---|---|---|---|
| ScW94 [11] | $^{241}$Pu($n_{th}$,f) | S I | 52.15±0.14 | 1.15±0.1 | 36.5±7.5 | Experiment (Values are given in *Z*) |
| | | S II | 54.92±0.37 | 1.91±0.16 | 63.5±6.3 | |

| Ref. | Nucleus | Channel | Position | Width | Yield (%) | Remarks |
|---|---|---|---|---|---|---|
| FaH95 [27] | $^{235}$U($n_{th}$,f) | S II | 140.6 | 4.53 | 82.1 | Experiment (See ref [27] for more energies.) |
| | | S I | 133.1 | 2.47 | 17.8 | |
| | | SL | 118.0 | 4.88 | 0.1 | |
| | $^{235}$U(n,f) $E_n$ = 6 MeV | S II | 140.6 | 4.93 | 78.4 | |
| | | S I | 133.1 | 2.71 | 19.5 | |
| | | SL | 118.0 | 5.27 | 2.1 | |

| Ref. | Nucleus | Channel | Position | Width | Yield (%) | Remarks |
|---|---|---|---|---|---|---|
| SiH95 [13] | $^{237}$Np(n,f) $E_n$=5.5MeV | S I | 134.7±0.1 | 3.7±0.1 | - | Experiment |
| | | S II | 140.3±0.1 | 6.4±0.1 | - | |
| | | S III | (153.0±3.0) | (4.9±1.2) | - | |
| | | SL | (119.0±3.0) | (13.5±3.4) | - | |
| | $^{238}$Np | S I | 135.4±3.0 | 3.9±1.0 | - | Calculated |
| | | S II | 139.1±3.0 | 5.9±1.5 | - | |
| | | S III | 153.0±3.0 | 4.9±1.2 | - | |
| | | SL | 119.0±3.0 | 13.5±3.4 | - | |

| Ref. | Nucleus | Channel | Position | Width | Yield (%) | Remarks |
|---|---|---|---|---|---|---|
| WaD96 [28] | $^{243}$Am($n_{th}$,f) | S I | - | - | 38 | Experiment |
| | | S II | - | - | 62 | |

| Ref. | Nucleus | Channel | Position | Width | Yield (%) | Remarks |
|---|---|---|---|---|---|---|
| DeW97 [12] | $^{236}$Pu(s.f.) | S I | 132.0±0.3 | 2.7±0.4 | 9.7±2.1 | Experiment |
| | | S II | 140.3±0.2 | 5.2±0.1 | 90.3±2.1 | |
| | $^{238}$Pu(s.f.) | S I | 133.5±0.2 | 1.6±0.2 | 6.3±1.4 | |
| | | S II | 140.1±0.2 | 5.8±0.2 | 93.7±1.4 | |
| | $^{240}$Pu(s.f.) | S I | 134.60±0.06 | 2.67±0.06 | 26.4±1.1 | |
| | | S II | 140.37±0.10 | 5.73±0.05 | 73.6±1.1 | |
| | $^{242}$Pu(s.f.) | S I | 134.68±0.03 | 2.69±0.02 | 37.3±0.5 | |
| | | S II | 139.65±0.05 | 5.34±0.03 | 62.7±0.5 | |
| | $^{244}$Pu(s.f.) | S I | 134.91±0.05 | 3.41±0.04 | 44.5±1.1 | |
| | | S II | 140.88±0.13 | 5.85±0.05 | 55.5±1.1 | |

| Ref. | Nucleus | Channel | Position | Width | Yield (%) | Remarks |
|---|---|---|---|---|---|---|
| ObH98 [29] | $^{239}$U | S I | 137±3 | 4.2±1.0 | - | Calculated |
| | | S X | 139±3 | 4.8±1.0 | - | |
| | | S II | 142±3 | 4.5±1.0 | - | |
| | | SL | 120±3 | 8.7±1.0 | - | |
| | $^{238}$U(n,f) $E_n$=5.5 MeV | S I | 135.2±0.1 | 3.4±0.1 | - | Experiment |
| | | S II | 141.3±0.1 | 6.2±0.1 | - | |
| | | SL | (119.5±3.0) | (13±3.4) | - | |

| Ref. | Nucleus | Channel | Position | Width | Yield (%) | Remarks |
|---|---|---|---|---|---|---|
| BrK99 [15] | $^{235}$U(n,f) E*=6.55 MeV | S I | 100.10 | 2.93 | 24.5±0.8 | Experiment (See [15] for original work and more energies.) |
| | | S II | 94.78 | 4.86 | 75.4±0.8 | |
| | | SL | 118 | 4.88 | 0.071±0.013 | |
| | $^{235}$U(n,f) E*=12.55MeV | S I | - | - | 18.4±0.5 | |
| | | S II | - | - | 80.8±2.1 | |
| | | SL | - | - | 0.79±0.2 | |
| | $^{232}$Th(γ,f) E*=6.44 MeV | S I | - | - | 19.9±2.0 | |
| | | S II | - | - | 80.1±2.9 | |
| | | SL | - | - | 0.01±0.06 | |
| | $^{232}$Th(γ,f) E*=13.15MeV | S I | - | - | 26.8±1.1 | |
| | | S II | - | - | 72.2±1.3 | |
| | | SL | - | - | 1.01±0.2 | |
| | $^{237}$Np(n,f) E*=5.99 MeV | S I | - | - | 32.4±1.4 | |
| | | S II | - | - | 67.5±3.1 | |
| | | SL | - | - | 0.17±0.13 | |
| | $^{237}$Np(n,f) E*=10.99MeV | S I | - | - | 22.4±1.6 | |
| | | S II | - | - | 76.7±2.6 | |
| | | SL | - | - | 0.81±0.13 | |

| Ref. | Nucleus | Channel | Position | Width | Yield (%) | Remarks |
|---|---|---|---|---|---|---|
| MuO99 [16] | $^{233}$Pa | S I | 133.8 | - | - | Experiment (Fission induced by 10.3 MeV protons.) |
| | | S II | 139.5 | - | - | |
| | | S III | 148.8 | - | - | |
| | | SL | 84.2 | - | - | |
| | $^{234}$Np | S I | 133.0 | - | - | |
| | | S II | 138.0 | - | - | |
| | | S III | 149.8 | - | - | |
| | | SL | 84.2 | - | - | |
| | $^{236}$Np | S I | 133.9 | - | - | |
| | | S II | 139.0 | - | - | |
| | | S III | 153.3 | - | - | |
| | | SL | 82.7 | - | - | |
| | $^{237}$Np | S I | 134.0 | - | - | |
| | | S II | 139.4 | - | - | |
| | | S III | 153.7 | - | - | |
| | | SL | 83.3 | - | - | |
| | $^{239}$Np | S I | 134.6 | - | - | |
| | | S II | 140.2 | - | - | |
| | | S III | 155.6 | - | - | |

| | | | | |
|---|---|---|---|---|
| | SL | 83.4 | - | - |
| $^{239}$Am | S I | 133.8 | - | - |
| | S II | 139.1 | - | - |
| | S III | 155.0 | - | - |
| | SL | 84.0 | - | - |
| $^{240}$Am | S I | 134.5 | - | - |
| | S II | 139.8 | - | - |
| | S III | 156.5 | - | - |
| | SL | 83.6 | - | - |
| $^{241}$Am | S I | 134.7 | - | - |
| | S II | 140.3 | - | - |
| | S III | 156.4 | - | - |
| | SL | 84.5 | - | - |
| $^{243}$Am | S I | 135.2 | - | - |
| | S II | 140.9 | - | - |
| | S III | 157.0 | - | - |
| | SL | 86.0 | - | - |
| $^{245}$Bk | S I | 135.7 | - | - |
| | S II | 141.7 | - | - |
| | S III | 158.6 | - | - |
| | SL | 86.4 | - | - |

| Ref. | Nucleus | Channel | Position | Width | Yield (%) | Remarks |
|---|---|---|---|---|---|---|
| ViH00 [30] | $^{238}$U(n,f) | S I | - | - | 41.56±1.0 | Experiment |
| | $E_n$=1.2MeV | S II | - | - | 58.44±1.0 | (See [30] for |
| | $^{238}$U(n,f) | S I | - | - | 25.57±0.067 | more ener- |
| | $E_n$=5.8MeV | S II | - | - | 73.1±0.344 | gies.) |

| Ref. | Nucleus | Channel | Position | Width | Yield (%) | Remarks |
|---|---|---|---|---|---|---|
| HaV00 [31] | $^{237}$Np(n,f) | S I | - | - | 33.00±0.02 | Experiment |
| | $E_n$=0.5MeV | S II | - | - | 66.92±0.02 | (See [31] for |
| | $^{237}$Np(n,f) | S I | - | - | 24.58±0.05 | more ener- |
| | $E_n$=5.5MeV | S II | - | - | 72.93±0.06 | gies.) |